\documentclass[journal]{IEEEtran}
\usepackage{url}
\usepackage{verbatim}
\usepackage{amsmath}
\usepackage{amssymb}
\usepackage{textcomp}
\usepackage{cite}
\usepackage{textpos}
\usepackage{graphicx}
\usepackage{longtable}
\usepackage{subfigure}
\usepackage{enumerate}
\usepackage{tabularx,booktabs}
\usepackage[flushleft]{threeparttable}
\usepackage{siunitx}
\usepackage{color}
\usepackage[utf8]{inputenc}
\usepackage{dirtytalk}
\graphicspath{ {./images/} }

\usepackage{multirow}

\usepackage{adjustbox}
\usepackage{textcomp}

\usepackage{algorithmic}
\usepackage[lined,boxed,ruled,commentsnumbered]{algorithm2e}
\usepackage{longtable}
\usepackage{subfiles}

\makeatletter
\newcommand*\bigcdot{\mathpalette\bigcdot@{1.5}}
\newcommand*\bigcdot@[2]{\mathbin{\vcenter{\hbox{\scalebox{#2}{$\m@th#1\bullet$}}}}}
\makeatother

\newcolumntype{M}[1]{>{\centering\arraybackslash}m{#1}}
\newcolumntype{C}{>{\centering\arraybackslash}X} 
\renewenvironment{IEEEbiography}[1]
  {\IEEEbiographynophoto{#1}}
  {\endIEEEbiographynophoto}

\begin{document}

\title{Towards the Age of Intelligent Vehicular Networks for Connected and Autonomous Vehicles in 6G}

\author{Van-Linh Nguyen, \textit{Member, IEEE}, Ren-Hung Hwang, \textit{Senior Member, IEEE}, Po-Ching Lin, \textit{Member, IEEE}, Abhishek Vyas, \textit{Member, IEEE}, Van-Tao Nguyen  \thanks{Van-Linh Nguyen, Ren-Hung Hwang, Po-Ching Lin, Abhishek Vyas are with National Chung Cheng University, Minhsiung, Chiayi, Taiwan.} \thanks{Van-Tao Nguyen are also with Thai Nguyen University of Information and Communication Technology, Thai Nguyen, Vietnam.}
}

\markboth{ }
{Van-Linh Nguyen \MakeLowercase{\textit{et al.}}: Towards the Age of Intelligent Vehicular Networks for Connected and Autonomous Vehicles in 6G}

\maketitle

\begin{abstract}
 
Twenty-two years after the advent of the first-generation vehicular network, i.e., dedicated short-range communications (DSRC) standard/IEEE 802.11p, the vehicular technology market has become very competitive with a new player, Cellular Vehicle-to-Everything (C-V2X). Currently, C-V2X technology likely dominates the race because of the big advantages of comprehensive coverage and high throughput/reliability. Meanwhile, DSRC-based technologies are struggling to survive and rebound with many hopes betting on success of the second-generation standard, IEEE P802.11bd. While the standards battle to attract automotive makers and dominate the commercial market landing, the research community has started thinking about the shape of the next-generation vehicular networks. This article details the state-of-the-art progress of vehicular networks, particularly the cellular V2X-related technologies in specific use cases, compared to the features of the current generation. Through the typical examples, we also highlight why 5G is inadequate to provide the best connectivity for the vehicular applications and then 6G technologies can fill up the vacancy.

\end{abstract}


\IEEEpeerreviewmaketitle
  
 \section{Introduction}
 
 Vehicular communications are critical technologies to assist vehicles in areas with limited visibility where cameras and advanced techniques like radar operate poorly. Due to the importance of the technology, Federal Communications Commission (FCC) in United States allocated 75MHz \cite{FCC_Detail} of spectrum in the 5.9GHz WiFi band for Dedicated Short-Range Communications (DSRC), a technology specifically designed for automotive use, in 1999. The allocation brought hope to fuel many innovations for enhancing road safety and traffic efficiency. However, contrary to the expectation, DSRC has never been widely deployed. In November 2020, the FCC unanimously approved reallocating 45 MHz of the 5.9GHz spectrum to WiFi and other unlicensed uses citing the lack of adoption \cite{FCC_Detail}. A spectrum of 30MHz is still kept for transportation-related services but expected to transition to C-V2X eventually.  
	 
	 \begin{figure*}[ht]
		\begin{center}
			\includegraphics[width=1\linewidth]{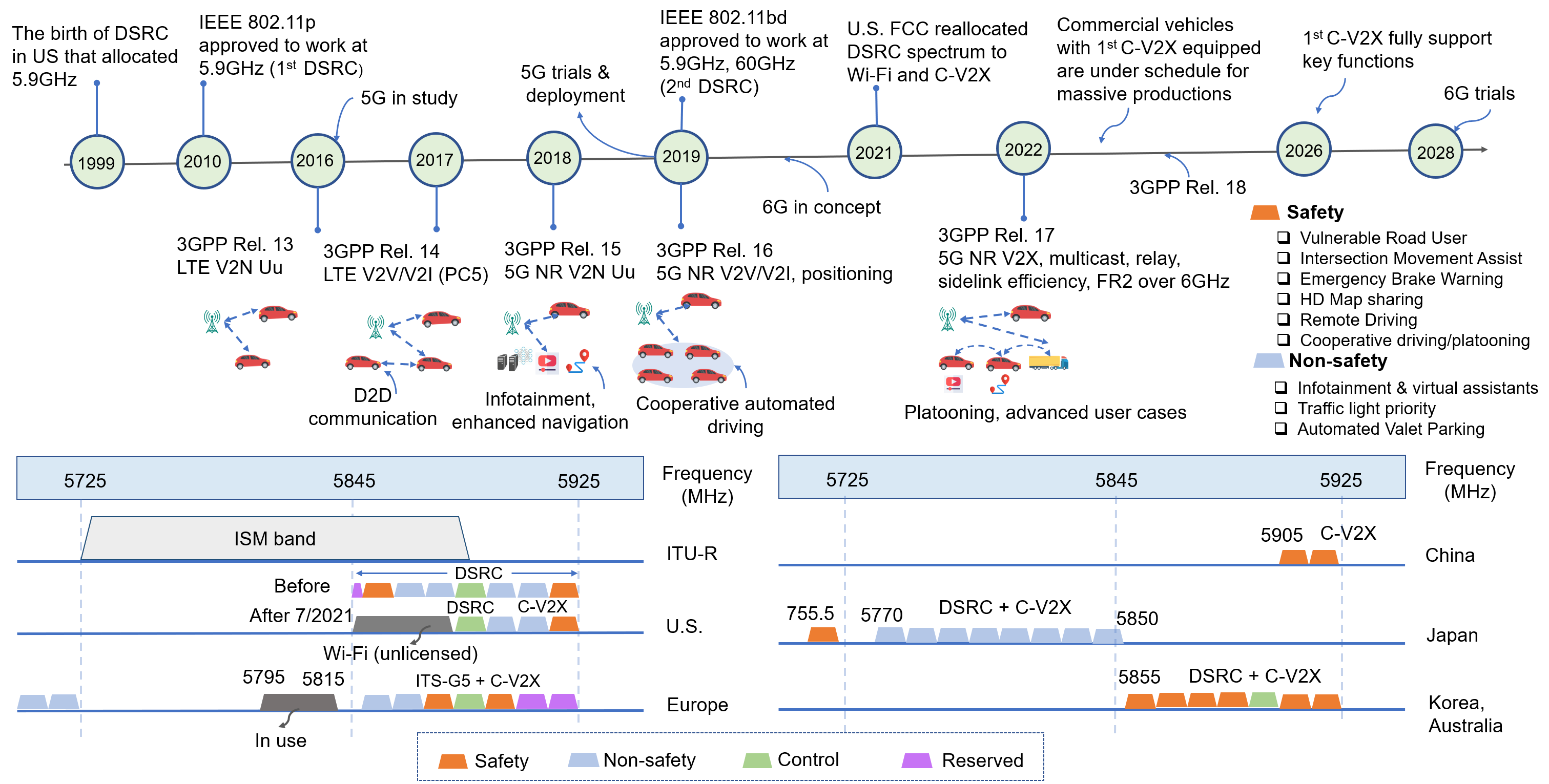}
		\end{center}
	\centering
	 	\caption{The roadmap of DSRC and C-V2X. After 7/2021, the DSRC frequency (in US) will be allocated to Wi-Fi and C-V2X. In China, Korea, the agency starts to move to the C-V2X standard. Europe and Japan maintain the technology-neutral spectrum designation regime. Some automakers are shifting to C-V2X (e.g., Audi, Nissan) while the others are still enthusiastic DSRC supporters (e.g., Volkswagen, Toyota). }
	 	\label{fig:DSRC-C-V2X-roadmap}
	 \end{figure*}

    Unlike DSRC, starting from the first version released in 2016 (under 3GPP Release 13), C-V2X technologies  grow rapidly and are likely ready for commercial deployment in 2022, a record-short research-to-deployment process. Figure~\ref{fig:DSRC-C-V2X-roadmap} summarizes key timelines for the development of DSRC and C-V2X. Generally, C-V2X can replace all DSRC features such as vehicle-to-vehicle (V2V) and Vehicle-to-Infrastructure(V2I) communications. Besides, by exploiting the cellular enterprise infrastructure, C-V2X can enable robust security protection, network reliability, and scalability at a level that DSRC cannot afford. Further, C-V2X offers new communication types such as Vehicle-to-Pedestrian (V2P) and Vehicle-to-Network (V2N). In short, the DSRC spectrum reallocation plan from FCC likely breaks the status quo and expands the market share for C-V2X. The future of DSRC will rely on the success of the second generation, IEEE P802.11bd, or beyond. A possible solution for the coexistence, which ETSI Technical Group 37 is exploring, is to allow frequency channel sharing in an area. For example, DSRC can be activated at the places where C-V2X signals are weak (rural/remote areas).

  Currently, the research community has started to move quickly on the subsequent phases of the next-generation vehicular network research. The following section presents our examples of future V2X applications and new key enhancements for C-V2X toward 6G era. The enhancements are then mapped with the use cases to highlight how 6G enabling technologies can satisfy the use cases' technical requirements. Finally, we discuss V2X open problems and conclude the paper.

 \begin{figure}[ht]
	\begin{center}
		\includegraphics[width=1\linewidth]{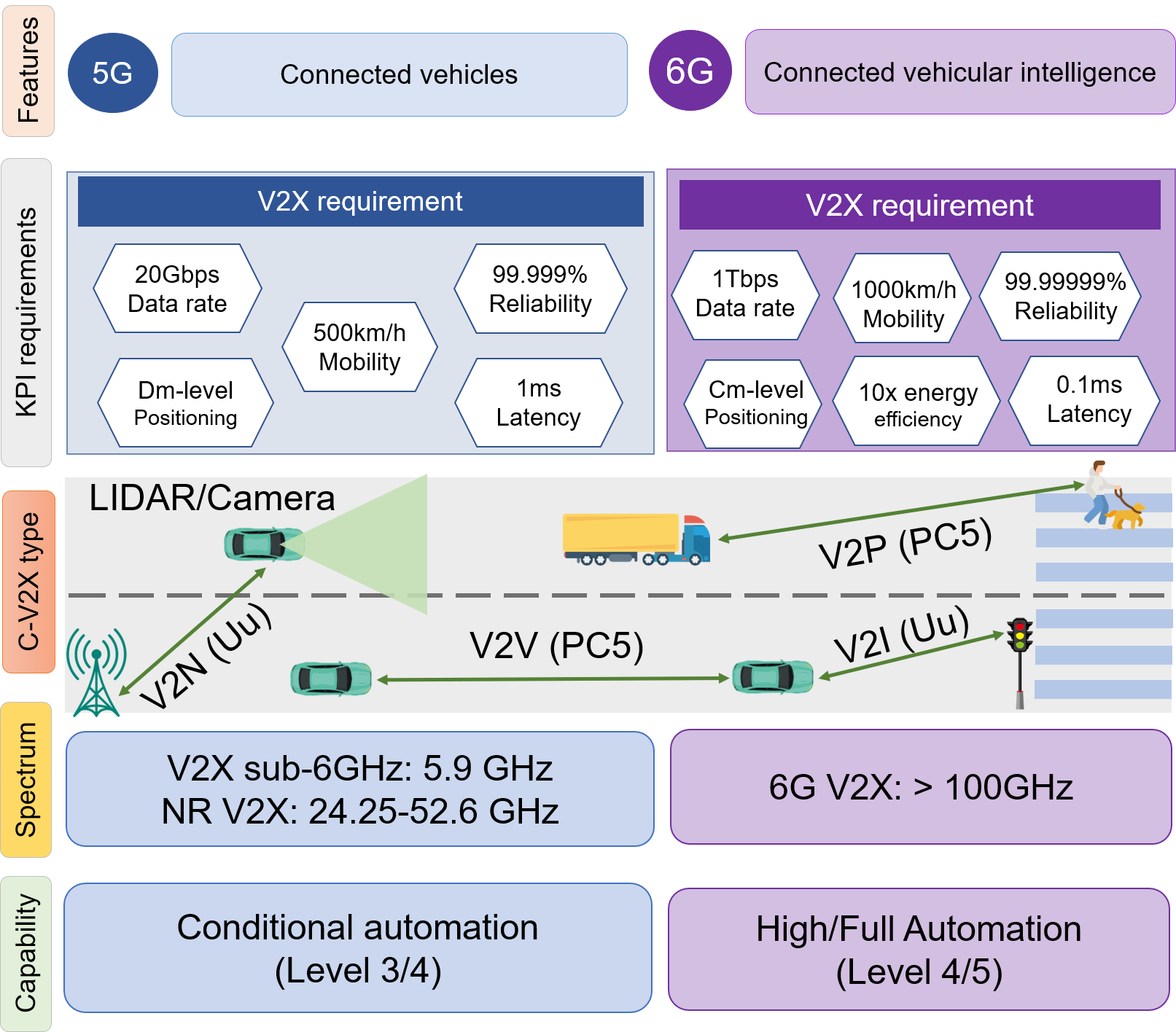}
	\end{center}
\centering
 	\caption{The taxonomy of V2X in the 5G and 6G vision in terms of network requirements, spectrum, and capability. Some acronyms are Light Detection and Ranging (LIDAR), New Radio (NR), vehicle-to-vehicle (V2V), Vehicle-to-Pedestrian (V2P), Vehicle-to-Network (V2N), and Vehicle-to-Infrastructure(V2I). Uu and PC5 are communication interfaces.}
 	\label{fig:5G-6G-roadmap}
 \end{figure}
 
 \begin{figure*}[ht]
		\begin{center}
			\includegraphics[width=1\linewidth]{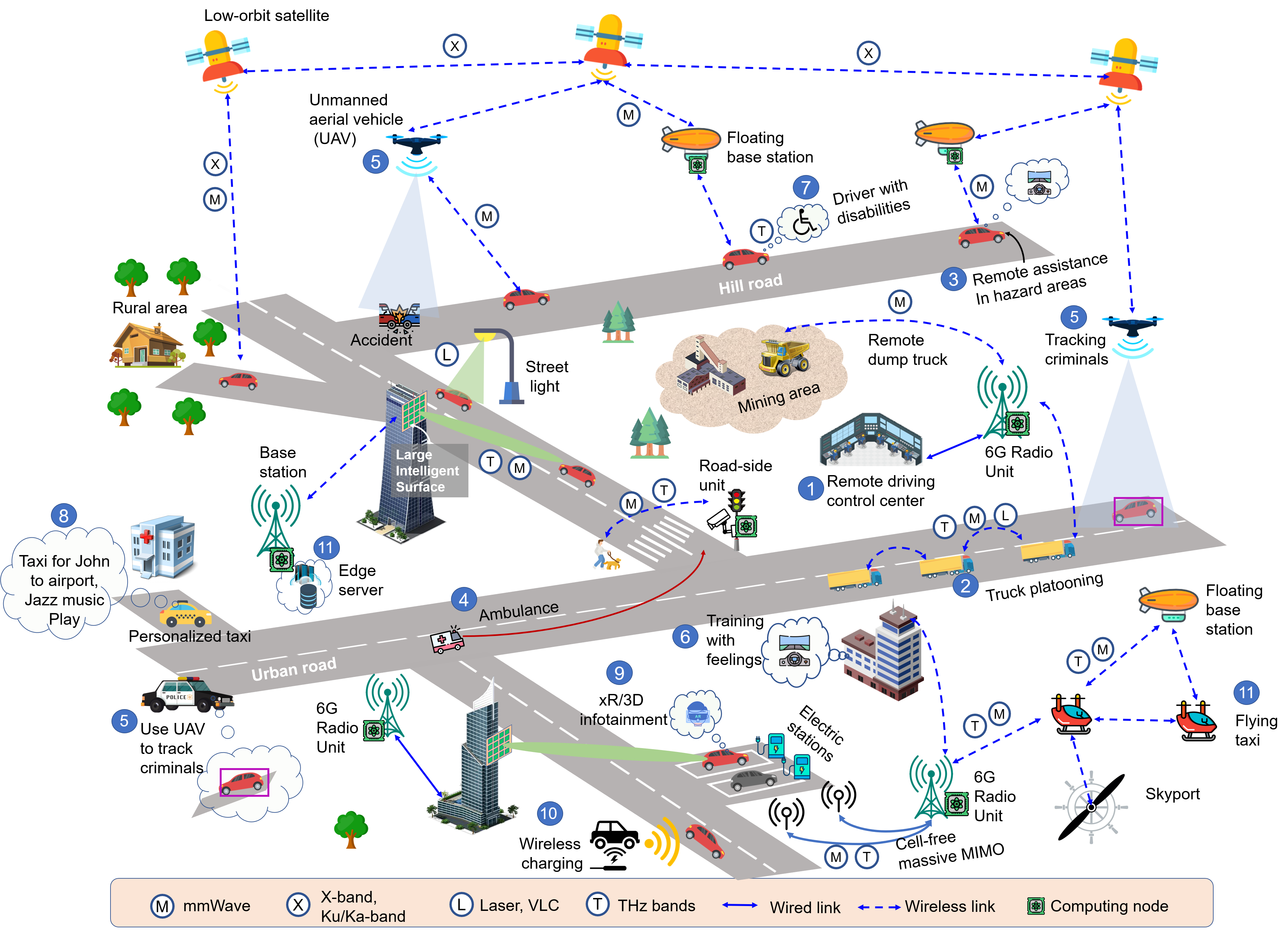}
		\end{center}
	\centering
	 	\caption{Vehicular communications in the vision of connected and autonomous vehicles (CAV) and eleven examples of using such technologies in five specific use cases. The frequency range of the communication technologies: mmWave(sub-6GHz: 2.4-5.9GHz, NR: 24.25-52.5GHz); X-band (8-12 GHz); Ku-band (12–18 GHz); Ka-band (27–40 GHz); Laser,VLC (400-800THz); THz(300GHz to 3THz). Extremely high-frequency spectrum for V2X ($>100GHz$) are still standardized. Different countries and territories may have separate regulations on frequency bands.}
	 	\label{fig:vehicular-communication-6G-networks}
	 \end{figure*}

\section{6G vision and novel V2X applications}

6G is the sixth generation standard for cellular communications that are currently in the concept discussion stage. 6G is scheduled to succeed 5G and commercially deploy someday in the 2030s. Connected and autonomous vehicles will be one of the most promising applications that are expected to fully operate in 6G \cite{Tang2020}. Figure~\ref{fig:5G-6G-roadmap} summarizes key differences between 5G and 6G for V2X in terms of features, Key Performance Indicator (KPI) requirements, spectrum usage, and vehicle capability. Generally, 6G is expected to be different from 5G on several key KPIs: data rate (50x faster, 1Tbps vs 20Gbps), service latency (10x lower, 0.1ms vs 1ms), positioning capability (10x more accurate, cm-level vs dm-level), reliability (100x more reliable), wider coverage, energy efficiency (10x better), and so on \cite{Saad2020}. Further, 6G can enable vehicular intelligence and high/full vehicle automation capability. Five following use cases illustrate how future V2X applications demand such network capabilities. The examples make a case to clarify why 5G cannot provide the best connectivity quality while 6G can.

\textbf{Use Case 1: Advanced remote driving and autonomous driving to deploy in urban areas with new business models.} 

Remote driving and autonomous vehicles are by no means state-of-the-art technology. Mining giant Rio Tinto in Western Australia has operated an operations centre 1,200 kilometres away from the field to control driverless trucks for carrying mined iron ore. In the control room, as illustrated in the case \textcircled{\raisebox{-0.9pt}{1}} of Figure~\ref{fig:vehicular-communication-6G-networks}, remote drivers work as office workers to interfere only when difficult road situations arise such as bad weather that hampers sensors and computers from working smoothly. Generally, communication technologies like 4G/5G are necessary to support remote interventions. 

However, remote driving in 6G will evolve even further in terms of new features and business coverage. As illustrated in the cases \textcircled{\raisebox{-0.9pt}{2}} and \textcircled{\raisebox{-0.9pt}{3}} of Figure~\ref{fig:vehicular-communication-6G-networks}, remote driving in 6G can operate in both urban and remote areas. Those can be remote assistance for civil clients when they drive in hazardous areas or truck platooning services. 5G cannot cover connectivity for such a remote area. Besides, unlike driving in a little-interference environment (e.g., mining sites), remote driving in urban areas will be a tremendous challenge because of unpredictable interference of human drivers. For safety control, transmitting high-quality video (holographic visuals) and multi-sensor data is often required to help remote drivers capture the rapid changes of traffic states around the vehicles. However, transmitting full-featured 8K/16K ultra high resolution or higher quality videos may require up to dozens of Gbps. This extremely high throughput is still challenging for current communication technologies (even 5G) to satisfy. 

\textbf{Use Case 2: Intelligent traffic scheduling and road-safety control to replace the current traffic light control or coordination by police officers}

With millions of people killed by road traffic collisions worldwide every year, traffic scheduling and road-safety applications are expected to be the top applications to be implemented in vehicular networks. In 6G, when vehicular networks become mature and widely deployed, the applications will support new features. For example, powerful AI can proactively coordinate in-out traffic (signal periods) through fully-connected traffic lights at the road intersections to enhance road capacity, reduce traffic jams or prioritize the lanes for emergency vehicles (the case \textcircled{\raisebox{-0.9pt}{4}} of Figure~\ref{fig:vehicular-communication-6G-networks}). With advantages of high reliability in 6G communications and intelligence, autonomous vehicles can also ``interact'' with each other to build an exact moving schedule and smoothly pass through an unsignalized intersection. Unmanned Aerial Vehicle (UAV) and flying base stations become new facilities to assist the tracking and traffic coordination effectively, particularly in the remote areas (the case \textcircled{\raisebox{-0.9pt}{5}} of Figure~\ref{fig:vehicular-communication-6G-networks}) where lack of 5G signals is common \cite{Hu21}. 
	 
\textbf{Use Case 3: Holographic driving and brain-controlled vehicles in the 6G era}

 Augmented reality (AR), virtual reality (VR), and mixed reality (MR) are 5G-driven technologies that have been partially deployed in practice. In 6G, they will be enhanced and combined into extended reality (XR) to enable metaverse VR or digital twin environments along with human-machine interactions. In this aspect, 6G V2X can assist remote drivers in connecting with vehicle sensors to build a holographic vision of the surrounding environments and enhance the driver's senses (the case \textcircled{\raisebox{-0.9pt}{6}} of Figure~\ref{fig:vehicular-communication-6G-networks}). Further, 6G V2X communications can support brain-controlled vehicle (BCV), a futuristic scenario, where people with disabilities can have a chance to control their vehicles independently (the case \textcircled{\raisebox{-0.9pt}{7}} of Figure~\ref{fig:vehicular-communication-6G-networks}). Although many automakers are promoting the vision of fully autonomous vehicles, the involvement of humans in this case still brings up the irreplaceable faith and driving feelings. BCV requires ultra-high data rate communications which current wireless communication generations (4G/5G) cannot meet. For example, a coarse estimation of the whole brain recording demand is about 100 Gbps\cite{Guo21}, the transmission of which is beyond the theoretical peak data rate of 5G. 

\textbf{Use Case 4: Personalized transport vehicles with holographic infotainment and tactile/haptic interactions}

Many studies predict that most new-sale vehicles in the next decade will be electrified (e.g., by battery-electric, hydrogen fuel cell). Besides environmental protection, the transformation to use electricity in transport vehicles brings up new advantages. Abundant energy and advanced processors/sensors in future vehicles can enable initiatives for personalized infotainment and autonomous health monitoring in cars. Ride-hailing services are then able to be personalized at the caller's demands (the case \textcircled{\raisebox{-0.9pt}{8}} of Figure~\ref{fig:vehicular-communication-6G-networks}). Holographic games and streaming services are also expected to provide passengers with more leisure riding experience (the case \textcircled{\raisebox{-0.9pt}{9}} of Figure~\ref{fig:vehicular-communication-6G-networks}). While 5G can certainly provide connectivity for these services, it can be a challenge to maintain stable connections in rural and remote areas or enable haptic features. 6G can ease that trouble by expanding the coverage through integrated space-air-ground networks \cite{Saad2020}. Besides, in the future, charging electric vehicles via wireless may be no longer an imagination (the case \textcircled{\raisebox{-0.9pt}{10}} of Figure~\ref{fig:vehicular-communication-6G-networks}). New wireless-based charging and energy trading models (e.g., using blockchain) through 6G dense networks promise to transform the current fuel charging\cite{Wang22}.

\textbf{Use Case 5: Flying taxis and cargo drones to transform the future transportation model.}

Booking a flying taxi is no longer an imagination in the coming years (the case \textcircled{\raisebox{-0.9pt}{11}} of Figure~\ref{fig:vehicular-communication-6G-networks}). Several major brands like Uber and Hyundai or many startups have already spent much on developing flying taxis, skyports, and aerial ride-sharing/cargo drone services. As a result, many new functions such as flying parking lot management, cooperative flying, flight sharing/revocation, or high-end infotainment require good Internet communications and even sharing data among the flying vehicles, base stations, and probably remote control centers. Currently, although satellite-based Internet providers (e.g., Starlink, OneWeb) can provide limited connectivity, there exists no technology (even 5G) to provide communication interoperablity across multiple network infrastructure (space/air/ground/marine).

\section{6G technologies to enable network capabilities for V2X use case applications} 

\begin{figure*}[ht]
		\begin{center}
			\includegraphics[width=1\linewidth]{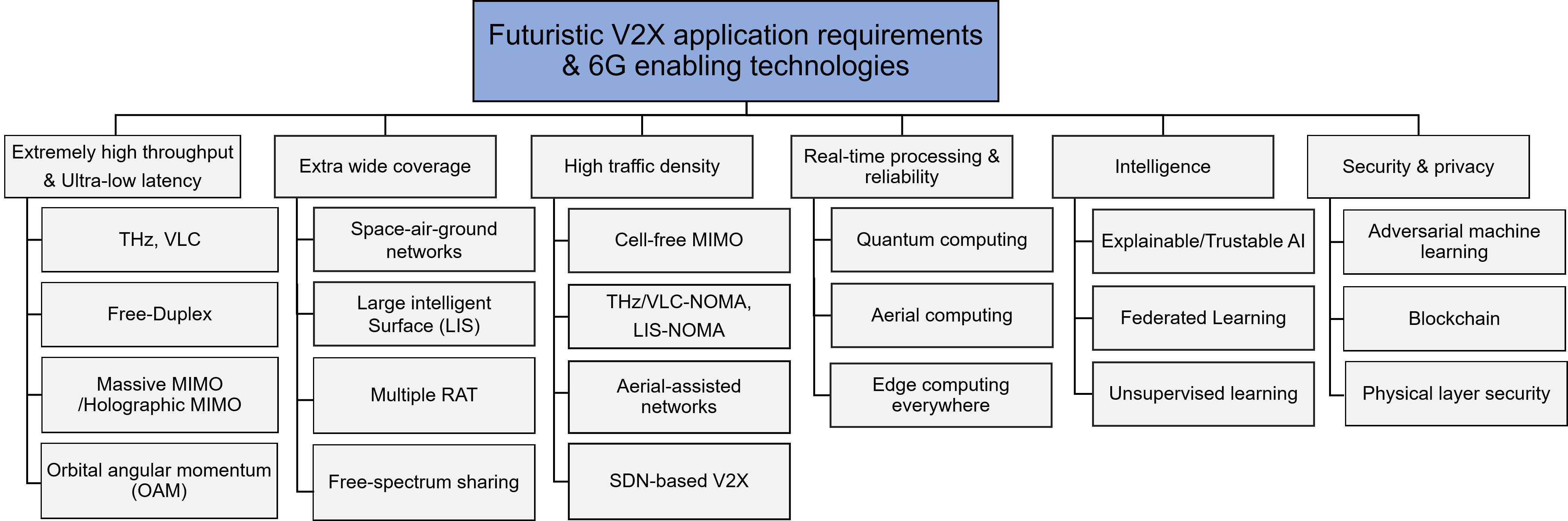}
		\end{center}
	\centering
	 	\caption{The taxonomy of several enabling technologies for 6G V2X by features. An application can require multiple features that are supported in different enabling technologies. Some acronyms are Terahertz (THz),  Visible Light Communication (VLC), Multiple-Input/Multiple-Out (MIMO), Large Intelligent Surface(LIS), Multiple Radio Access technology (Multi-RAT), Non-Orthogonal Multiple Access (NOMA), Artificial Intelligence (AI).}
	 	\label{fig:6G-V2X-technologies}
	 \end{figure*}

As mentioned earlier in five user cases, 5G is generally inadequate to gain the best experience or unlock full features for new V2X applications. To satisfy KPIs and realize the vision of these futuristic applications, current vehicular networks require new enhancements to many underlying technologies. Inspired by the vision in \cite{Saad2020,Tang2020,Guo21,Hu21,MEKRACHE2021100398,Manias21,Rahim22}, the taxonomy of the enhancements to appear in 6G-enabled vehicular networks is summarized in Figure~\ref{fig:6G-V2X-technologies}. Most technologies in the figure are still under development with various achievements, e.g., blockchain and massive MIMO. Some are still in the conceptual stage, e.g., Full/Free-Duplex and THz, due to many remaining technical challenges (overheat chip, antenna design). Several technologies in the list may fade away or not be ready at the time of 6G, e.g., edge computing everywhere and free-spectrum sharing. However, the research experience may provide valuable information for further breakthroughs. Generally, we filter the enabling technologies by specific requirements of the futuristic V2X applications in mentioned use cases and overview of how they work. Security and privacy issues are detailed in the next section.

\textbf{Enhancement 1: 6G enabling technologies for extremely high throughput and ultra low-latency V2X applications}

6G physical layer technologies such as THz communications and Visible Light Communication (VLC) \cite{Calvanese19} (as listed in the first column of Figure~\ref{fig:6G-V2X-technologies}) can provide super throughput of up to terabits per second and reliability that high-end V2X applications in Use Cases 1,2,4 demand. The technologies are parts of 6G ultra-reliable and low-latency communications (uRRLC) \cite{Calvanese19}. However, due to high absorption resonance, maintaining effective communications for high-mobility vehicles are then top concerns. Cell-free Multiple-Input and Multiple-Output (MIMO)/massive MIMO in 6G can be the solution \cite{Saad2020,Tang2020}. Cell-free MIMO comprises a large number of distributed, low cost, and low power access point antennas connected to a network controller. Since the cell-free MIMO system is not partitioned into cells and every user equipment (UE) can access the network simultaneously, the technology can significantly enhance mobile users' wireless transmission efficiency in urban areas.
  
Full degree of freedom duplex (Free-Duplex) is expected to be another breakthrough technology in 6G \cite{Tang2020}. By relaxing a requirement on Frequency Division Duplex (FDD)/Time Division Duplex (TDD) differentiation to full-duplex between the transceiver and transceiver links, Free-Duplex can unlock extremely high throughput and reduce transmission delay for low-latency V2X applications \cite{Saad2020}. Holographic radio are also key enhancements for communication paradigms in 6G. In V2X, they are particularly useful for short-range ($\sim$ 10m) V2X tactile applications (e.g., Use Case 3, 4). Holographic radio (a.k.a. holographic MIMO) \cite{Calvanese19} exploits the propagation characteristics offered by an electromagnetic channel with the support of three-dimensional (3D) beamforming or metal mirrors to enable 3D imaging, positioning, and communications in holographic infotainment or AR/VR applications (Use Case 3). Finally, Orbital Angular Momentum (OAM) - which uses electromagnetic wave characteristics associated with beam vorticity and phase singularity - is also tremendous potential for increasing the transmission capacity and spectral efficiency in  mmWave/THz-enabled vehicular networks \cite{Guo21}.

\textbf{Enhancement 2: 6G enabling technologies for extra wide coverage and vehicular traffic density applications}

 The excessive propagation loss and signal blockage in mmWave/THz/VLC advance the importance of using directional beamforming antennas. Nonetheless, the directional connectivity makes vehicular communications challenging for high-speed vehicles. In 6G, Large Intelligent Surfaces (LIS) promises to be an important technology to enhance the probability of getting directional line-of-sight communications \cite{Saad2020}. LIS is the integration of a massive number of antenna elements in a large low-cost panel installing on sky buildings (as illustrated in Figure~\ref{fig:vehicular-communication-6G-networks}) to act as an intelligent scatterer for mmWave/THz coverage improvement. Besides, as listed in the second and third column of Figure~\ref{fig:6G-V2X-technologies}, space-air-ground-sea integrated networks or aerial-assisted networks are the other key enabling technologies to expand 6G coverage \cite{Tang2020} in Use Case 1, 2, and 5. In essence, the networks use high-altitude platforms (e.g., UAV, flying stations) and low-earth-orbit satellites to relay signals and enhance line-of-sight transmission. 6G thus can enlarge the interconnectivity for V2X services beyond the territories where conventional vehicular networks cannot afford, e.g., remote driving in remote areas or flying taxi. Multiple radio access technology (multi-RAT), which supports a device to connect to multiple cellular networks (e.g., 5G, 6G) \cite{Rahim22}, is also useful for enhancing network coverage for V2X applications when the old generation networks have not yet fully been phased out. 
 
Non-Orthogonal Multiple Access (NOMA) is another promising technology to increase spectral efficiency and throughput in 6G V2X density applications \cite{Tang2020}. NOMA allows one frequency channel to serve multiple users within the same cell through utilizing superposition coding (SC) at the transmitter and successive interference cancellation (SIC) at the receiver. Besides, Software-defined networking (SDN)-based V2X models are also under active development for supporting scalability and flexibility of network configuration in traffic remote control (Use case 2) \cite{Tang2020}. Besides, traditionally, the lack of a matured technology for efficient spectrum-sharing and the constraints of spectrum regulations often force the operators to compete in building many base stations to cover many areas as possible but cannot use the others' spectrum resources or must do through a complicated roaming. Blockchain and AI-assisted autonomous systems to manage spectrum sharing and usage can enable a free-sharing spectrum vision and significantly improve the spectrum efficiency and coverage for multiple applications \cite{Wang22}, including V2X.

\textbf{Enhancement 3: High-performance computing for offloading assist in V2X delay-sensitive applications}

Latency is the enemy of many V2X applications. At the explosion of data from holographic infotainment and traffic scheduling in Use Cases 2, 3, and 4, new computing technologies in 6G (e.g., aerial/quantum computing as listed in the fourth column of Figure~\ref{fig:6G-V2X-technologies}) are key factors for accelerating data processing and offloading capability. However, due to the cost, quantum computing models may be not ready to apply widely at the early stage of 6G \cite{Saad2020}. The more feasible technology is the edge computing. Edge computing will likely be equipped in every hop of 6G communications \cite{Qi21} (edge-computing-everywhere), e.g., in UAVs, flying stations as illustrated in Figure~\ref{fig:vehicular-communication-6G-networks}. The edge-computing-everywhere model can assist real time off-loading tasks in a large number of connected and autonomous vehicles.

\textbf{Enhancement 4: Artificial intelligence (AI) to enable intelligent vehicular networks}

\begin{figure*}[ht]
		\begin{center}
			\includegraphics[width=1\linewidth]{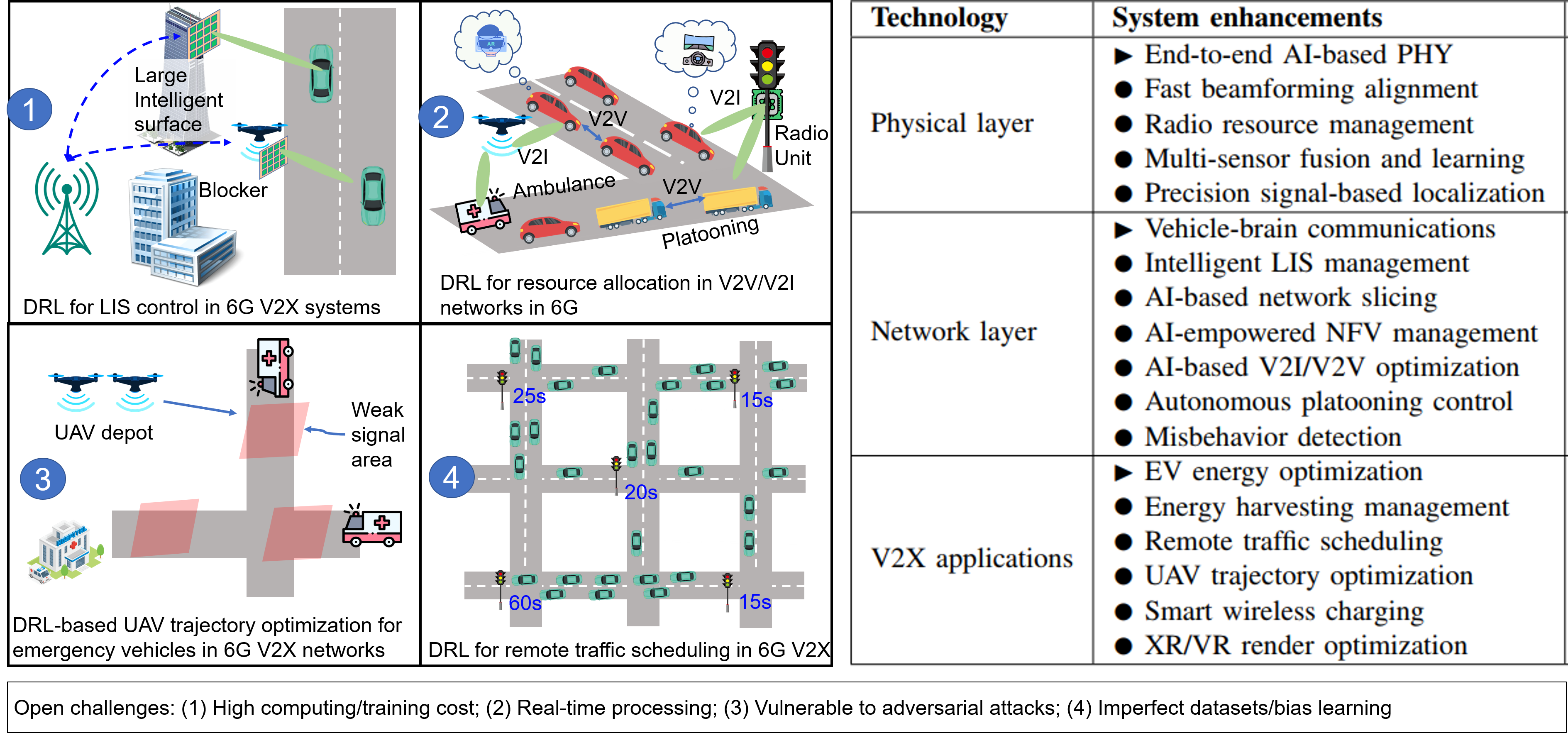}
		\end{center}
	\centering
	 	\caption{Illustration of four typical examples of using AI/machine learning models (e.g., Deep Reinforcement Learning (DRL)) to enable intelligent vehicular network functions (\textcircled{\raisebox{-0.9pt}{1}}, \textcircled{\raisebox{-0.9pt}{2}}) and V2X applications (\textcircled{\raisebox{-0.9pt}{3}}, \textcircled{\raisebox{-0.9pt}{4}}). Some other potential enhancements are listed on the right side.}
	 	\label{fig:AI-for-6G-V2X}
	 \end{figure*}

Intelligence will be a key difference between 5G V2X and 6G V2X. In 6G vehicular networks, vehicular intelligence (powered by machine/deep learning techniques) is a significant upgrade compared with the prior generations \cite{SHEHZAD22}. AI can help in two perspectives: (1) increase the proactive capability of traffic coordination and scheduling in the cities or high/full automation for vehicles (Use Case 1, 2, 5); (2) enhance accuracy performance, e.g., XR/VR renders in Use Case 3. Figure~\ref{fig:AI-for-6G-V2X} illustrates four typical examples of using AI technologies for enhancing vehicular networks. For example, the authors in \cite{Liu22} propose to use a deep reinforcement learning (DRL) to select the optimal phase shifts of reflecting surfaces for LIS-assisted vehicular networks (the case \textcircled{\raisebox{-0.9pt}{1}} of Figure~\ref{fig:AI-for-6G-V2X}). 

By training hundreds of thousands of the state-action examples (e.g., sub-channel and power level selections, selected channels patterns of neighbors, the time used for transmission, etc), the Dueling Deep Q Networks (DDQN)-based learning agent on each vehicle can figure out the optimal resource allocation policy for transmission \cite{MEKRACHE2021100398}. The intelligent resource allocation control is extremely meaningful for V2X dense/high-throughput applications (the case \textcircled{\raisebox{-0.9pt}{2}} of Figure~\ref{fig:AI-for-6G-V2X}). DRL-based AI systems are also powerful to find the optimal trajectory for multiple UAVs in providing connectivity for emergency rescue missions where ambulance vehicles may move through many weak signal areas (the case \textcircled{\raisebox{-0.9pt}{3}} of Figure~\ref{fig:AI-for-6G-V2X})). For large cities with complicated traffic conditions, DRL-based AI models can assist remote traffic centers in adjusting traffic light cycle flexibly and thus reducing traffic jams/waiting time in intersections (the case \textcircled{\raisebox{-0.9pt}{4}} of Figure~\ref{fig:AI-for-6G-V2X}). Some other potentials of using AI in different layers of V2X are summarized in the left side of Figure~\ref{fig:AI-for-6G-V2X}. 

    However, AI-empowered technologies in 6G certainly need more breakthroughs to gain high automation and reliability. The first target is to enhance the generative learning capability, e.g., vehicles or traffic control can auto-learn from the environment and operate correctly on previously unseen inputs. In this direction, meta-learning, and the combination of DRL and Generative Adversarial Network (GAN) are emerging approaches, given their strength in learning from massive inputs and automatically optimizing the decisions in interactive environments \cite{MEKRACHE2021100398}. Second, in the future, the personal infotainment in the vehicles will highly be personalized at the owner's demand \cite{Hui21}. To support such a feature, AI models need to collect private and sometimes sensitive information for learning. Federated learning (FL) can be the best suitable model for this kind of application to avoid privacy issues \cite{Liu21, Manias21}. FL coordinates the learning algorithms running in individual vehicles to improve the quality of a global learning model (at edge servers) without exchanging personal data. 

    To reduce the complexity of data collection, training, learning, end-to-end unsupervised or continual learning models like autoencoders can be a game-changer \cite{SHEHZAD22}. Autoencoder-based AI systems can learn and predict results directly on capturing online network traffic without labeling. Besides, the uncertainty of how an AI model works under hazardous environments has been a concern for years to apply AI for road-safety applications. To avoid the danger of AI making wrong decisions or biased learning, e.g., classifying wrong traffic signals, explainable AI models have gotten much attention recently. By proposing rule lists to study the DL decision-making process in specific contexts, Trustable/Explainable AI can provide an interpretable and faithful manner to the beam alignment decisions or traffic scheduling \cite{Guo21}.

\section{Discussion on open problems to realize intelligent vehicular networks in 6G}

Research for intelligent vehicular networks in 6G is at the early stage only. This section presents our discussion on several open problems for further research.

\textbf{Problem 1: The slow progress of V2X deployment and the lack of unified standards for many V2X technologies remain an issue.} 

Currently, the first generation of C-V2X is still stuck at trials and deployment. Neither LTE V2X nor 5G NR V2X has been evaluated as wide a range as DSRC did. V2X technologies for flying vehicles (Use Case 5) and Internet interroperability in space-air-ground networks remains open issues. Some critical features such as V2X authentication have not yet been standardized, although the US Department of Transportation (DOT) has just approved a Security Credential Management System (SCMS) proof-of-concept \cite{Tang2020}. However, it is unclear SCMS will be compatible with 5G V2X and how it looks in 6G. Moreover, the lack of consistent standards in the road, intelligent transportation system (ITS) architecture and legislation over countries is a grave concern. We believe that C-V2X and ITS will operate as an ecosystem. To gain a thriving ecosystem globally, numerous efforts and robust engagements are a must from industry automakers, developers, network operators, and transport authorities.

\textbf{Problem 2: Many technologies of vehicular communications, networking, computing, and intelligence have not yet been optimized.} 

6G-enabled vehicular technologies are at the infancy of development cycle, e.g., holographic radio. Conducting such technologies in specific use cases and then suggesting the optimal solutions (setting/placement) are the center of ongoing research efforts. Besides, energy efficiency is the top goal of many 6G-enabled V2X technologies. THz, tactile communication, slicing, or even edge/quantum computing all contribute to mitigating latency significantly in 6G but are simultaneously high energy consumption agents \cite{Rahim22}. Therefore, energy-friendly solutions such as dynamic service placement to gain substantial energy usage as 6G goals (10-100 times better than 5G \cite{Saad2020}), will be the top topics for future studies.

\textbf{Problem 3: Practical applications need to meet affordable billing models in C-V2X to be accepted widely}. 

Unlike DSRC, C-V2X communications in some ways require the involvement of the cellular operators. Therefore, fee billing for C-V2X infrastructure usage is likely the obstacle to getting users' acceptance of the business. The initiatives with affordable fee billing or phone bill merging models are thus important factors. This matter is urgent, but few studies have taken it seriously. A robust charging model for end-users will encourage V2X usage and generate revenue. Besides, we believe that the benefits from a rich set of new useful applications (not limited to the five use cases above) are the main factors to attract users and provide the merits for their upgrade decision to V2X.

\textbf{Problem 4: Security and privacy issues are the top concerns of current and future V2X technologies}.

\begin{figure}[ht]
		\begin{center}
			\includegraphics[width=1\columnwidth]{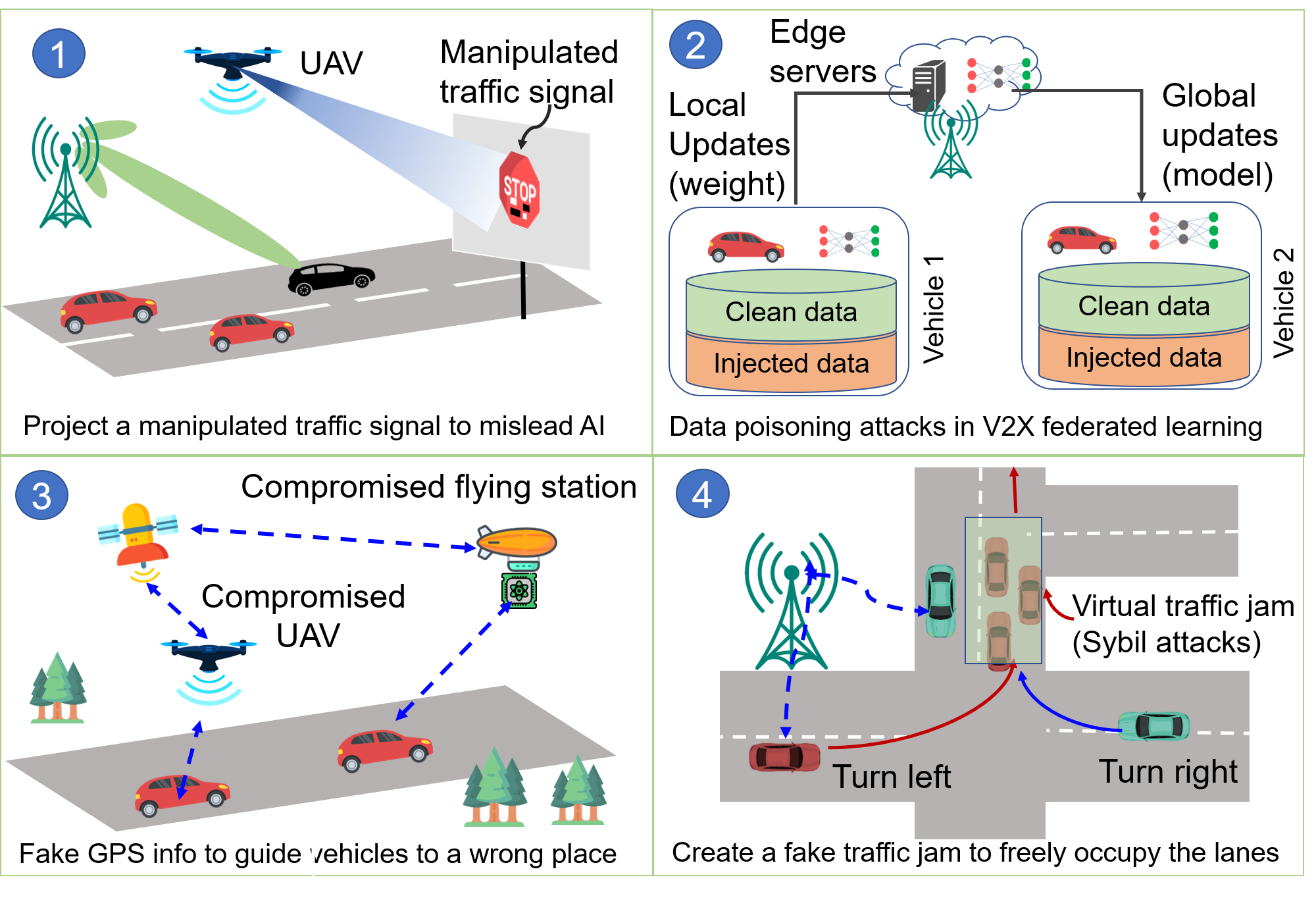}
		\end{center}
	\centering
	 	\caption{An illustration of four typical security attacks in 6G V2X: \textcircled{\raisebox{-0.9pt}{1}} Adversarial attacks; \textcircled{\raisebox{-0.9pt}{2}} Data poisoning;  \textcircled{\raisebox{-0.9pt}{3}} Compromised UAV;  \textcircled{\raisebox{-0.9pt}{4}} Sybil attacks.}
	 	\label{fig:intelligent-in-6G-V2X}
	 \end{figure}

The risk from security attacks has been the top concern for V2X commercial deployment. Unlike conventional C-V2X, AI will likely involve the process of many components of 6G V2X. AI can significantly enhance the automation and performance capacity of V2X applications but simultaneously the favorite targets of adversarial attacks. The adversarial attack means an attacker intentionally contaminates the input collection (inject incorrectly labeled data), algorithm (using learning weights designed by attackers), or training models (use an alternative model to replace the deployed model) to mislead vehicles' AI systems. Figure~\ref{fig:intelligent-in-6G-V2X} illustrates four typical examples of AI-targeted security attacks. In the case \textcircled{\raisebox{-0.9pt}{1}}, an attacker uses a UAV to project a 'modified STOP' traffic sign on a road banner to mislead the vehicles' AI-based detection applications that the vehicles are allowed to drive. As a result, the cars may accelerate and then be out of control due to sudden bad roads. In the case \textcircled{\raisebox{-0.9pt}{2}}, the attacker can manipulate the behavior of federated learning-based V2X applications by contaminating local updates (weights). If AI-based programs control the key components of intelligent vehicular networks, successful attacks can cause devastating damage, creating chaos in massive false alarms and further fatal accidents. Besides, the expansion of 6G V2X to non-territorial networks causes extra security concerns. A common case is an attacker uses compromised UAVs to falsify sharing information and guide vehicles to the attacker-designed places (case \textcircled{\raisebox{-0.9pt}{3}} in Figure~\ref{fig:intelligent-in-6G-V2X}). Various conventional threats in current V2X networks may remain. A prominent example is the attacker can use counterfeiting identities (pseudonyms) to create a fake traffic jam (e.g., by producing ghost vehicles on the high-definition map) and freely occupy the lanes (case \textcircled{\raisebox{-0.9pt}{4}} in Figure~\ref{fig:intelligent-in-6G-V2X}). 

In essence, data verification, physical layer security, and misbehavior detection are the common defense methods. For example, an AI designer can launch data sanitization (remove contaminated samples from training data) before performing an ML training process. The distributed data storage and mutual verification model in blockchain can help to single out unreliable data sources. For conventional threats, physical layer security (PLS) \cite{Guo21} (as listed in Figure~\ref{fig:6G-V2X-technologies}) is expected to enhance the trust of the vehicles and significantly prevent Sybil/tampering attacks. Given the immaturity of the technologies, further security studies on 6G V2X communications (e.g., THz-based V2X security, tactile communications), adversarial machine learning, and misbehavior detection will be essential topics in the coming years.

\section{Conclusion}
\label{sec:conclusion}

Vehicular networks are critical for future vehicles, particularly in obscured areas where the cameras and sensors operate poorly. This article details the state-of-the-art progress of the current vehicular communication standards and a glance about 6G V2X. The future V2X networks are expected to provide great connectivity for advanced features such as holographic rendering services and remote driving, even in remote areas. Based on the examples of five use cases, we have underscored 6G enabling technologies such as THz/VLC communications are critical to enhancing performance for future vehicular networks. Finally, we outline several open problems to realize the next-generation V2X. Accordingly, the consistency of V2X standards and ITS infrastructure, creative subscription model for C-V2X usage, and security/privacy issues are top issues. Meanwhile, optimizing 6G technologies for V2X, e.g., THz-NOMA, aerial-assisted vehicular networks will be the top promising techniques and important topics for further research.

\bibliographystyle{IEEEtran}
\bibliography{References.bib}

\begin{thebibliography}{10}
\providecommand{\url}[1]{#1}
\csname url@samestyle\endcsname
\providecommand{\newblock}{\relax}
\providecommand{\bibinfo}[2]{#2}
\providecommand{\BIBentrySTDinterwordspacing}{\spaceskip=0pt\relax}
\providecommand{\BIBentryALTinterwordstretchfactor}{4}
\providecommand{\BIBentryALTinterwordspacing}{\spaceskip=\fontdimen2\font plus
\BIBentryALTinterwordstretchfactor\fontdimen3\font minus
  \fontdimen4\font\relax}
\providecommand{\BIBforeignlanguage}[2]{{%
\expandafter\ifx\csname l@#1\endcsname\relax
\typeout{** WARNING: IEEEtran.bst: No hyphenation pattern has been}%
\typeout{** loaded for the language `#1'. Using the pattern for}%
\typeout{** the default language instead.}%
\else
\language=\csname l@#1\endcsname
\fi
#2}}
\providecommand{\BIBdecl}{\relax}
\BIBdecl

\bibitem{FCC_Detail}
{Federal Communications Commission}, ``Use of the 5.850-5.925 ghz band,''
  \emph{https://docs.fcc.gov/public/attachments/FCC-20-164A1.pdf}, accessed on
  10 June 2022.

\bibitem{Tang2020}
F.~Tang, Y.~Kawamoto, N.~Kato, and J.~Liu, ``Future intelligent and secure
  vehicular network toward {6G}: Machine-learning approaches,''
  \emph{Proceeding of IEEE}, vol. 108, no.~02, 2020.

\bibitem{Saad2020}
W.~Saad, M.~Bennis, and M.~Chen, ``A vision of {6G} wireless systems:
  Applications, trends, technologies, and open research problems,'' \emph{IEEE
  Network}, vol.~34, no.~3, pp. 134--142, 2020.

\bibitem{Hu21}
J.~Hu, C.~Chen, L.~Cai, M.~R. Khosravi, Q.~Pei, and S.~Wan, ``{UAV}-assisted
  vehicular edge computing for the {6G} internet of vehicles: Architecture,
  intelligence, and challenges,'' \emph{IEEE Communications Standards
  Magazine}, vol.~5, no.~2, pp. 12--18, 2021.

\bibitem{Guo21}
H.~Guo, X.~Zhou, J.~Liu, and Y.~Zhang, ``Vehicular intelligence in {6G}:
  Networking, communications, and computing,'' \emph{Vehicular Communications},
  vol.~33, p. 100399, 2022.

\bibitem{Wang22}
J.~Wang, K.~Zhu, and E.~Hossain, ``{Green Internet of Vehicles (IoV) in the 6G
  Era: Toward Sustainable Vehicular Communications and Networking},''
  \emph{IEEE Transactions on Green Communications and Networking}, vol.~6,
  no.~1, pp. 391--423, 2022.

\bibitem{MEKRACHE2021100398}
A.~Mekrache, A.~Bradai, E.~Moulay, and S.~Dawaliby, ``Deep reinforcement
  learning techniques for vehicular networks: Recent advances and future trends
  towards {6G},'' \emph{Vehicular Communications}, p. 100398, 2021.

\bibitem{Manias21}
D.~M. Manias and A.~Shami, ``Making a case for federated learning in the
  internet of vehicles and intelligent transportation systems,'' \emph{IEEE
  Network}, vol.~35, no.~3, pp. 88--94, 2021.

\bibitem{Rahim22}
M.~Noor-A-Rahim, Z.~Liu, H.~Lee, M.~O. Khyam, J.~He, D.~Pesch, K.~Moessner,
  W.~Saad, and H.~V. Poor, ``{6G for Vehicle-to-Everything (V2X)
  Communications}: Enabling technologies, challenges, and opportunities,''
  \emph{Proceedings of the IEEE}, vol. 110, no.~6, pp. 712--734, 2022.

\bibitem{Calvanese19}
E.~Calvanese~Strinati, S.~Barbarossa, J.~L. Gonzalez-Jimenez, D.~Ktenas,
  N.~Cassiau, L.~Maret, and C.~Dehos, ``{6G}: The next frontier: From
  holographic messaging to artificial intelligence using subterahertz and
  visible light communication,'' \emph{IEEE Vehicular Technology Magazine},
  vol.~14, no.~3, pp. 42--50, 2019.

\bibitem{Qi21}
W.~Qi, Q.~Li, Q.~Song, L.~Guo, and A.~Jamalipour, ``Extensive edge intelligence
  for future vehicular networks in {6G},'' \emph{IEEE Wireless Communications},
  pp. 1--8, 2021.

\bibitem{SHEHZAD22}
M.~K. Shehzad, L.~Rose, M.~M. Butt, I.~Z. Kovacs, M.~Assaad, and M.~Guizani,
  ``Artificial intelligence for {6G Networks}: Technology advancement and
  standardization,'' \emph{IEEE Vehicular Technology Magazine}, pp. 2--11,
  2022.

\bibitem{Liu22}
X.~Liu, Y.~Deng, C.~Han, and M.~D. Renzo, ``Learning-based prediction,
  rendering and transmission for interactive virtual reality in {RIS}-assisted
  terahertz networks,'' \emph{IEEE Journal on Selected Areas in
  Communications}, vol.~40, no.~2, pp. 710--724, 2022.

\bibitem{Hui21}
Y.~Hui, N.~Cheng, Z.~Su, Y.~Huang, P.~Zhao, T.~H. Luan, and C.~Li, ``Secure and
  personalized edge computing services in {6G} heterogeneous vehicular
  networks,'' \emph{IEEE Internet of Things Journal}, pp. 1--1, 2021.

\bibitem{Liu21}
S.~Liu, J.~Yu, X.~Deng, and S.~Wan, ``Fedcpf: An efficient-communication
  federated learning approach for vehicular edge computing in {6G}
  communication networks,'' \emph{IEEE Transactions on Intelligent
  Transportation Systems}, pp. 1--14, 2021.

\end{thebibliography}

   \begin{IEEEbiography}{Van-Linh Nguyen} (M'19) is an assistant professor at the Department of Computer Science and Information Engineering, National Chung Cheng University (CCU), Taiwan, and head of the Cybersecurity Lab. His research interests include cybersecurity, physical layer security, vehicular networks, and autonomous driving.
	\end{IEEEbiography}
	
	\begin{IEEEbiography}{Ren-Hung Hwang} (M'87) is the Dean of the College of Artificial Intelligence, National Yang Ming Chiao Tung University (NYCU), Taiwan. Before joining NYCU, he was with National Chung Cheng University, Taiwan, from 1993 to 2022. His current research interest is in Deep Learning, Wireless Communications, Network Security, and Cloud/Edge/Fog Computing.
   \end{IEEEbiography}
	
	\begin{IEEEbiography}{Po-Ching Lin} (M'09) is a professor at the Department of Computer Science and Information Engineering, National Chung Cheng University (CCU), Taiwan. His research interests include network security, network traffic analysis, and performance evaluation of network systems. 
	\end{IEEEbiography}
	
	\begin{IEEEbiography}{Abhishek Vyas} is a Ph.D. candidate at the Department of Computer Science and Information Engineering, National Chung Cheng University (CCU), Taiwan. His research interests include Cybersecurity, Computer Networks, and Non Terrestrial Networks. 
	\end{IEEEbiography}
	
	\begin{IEEEbiography}{Van-Tao Nguyen} is the president of Thai Nguyen University of Information and Communication Technology, Thai Nguyen, Vietnam. His research interests include Data Security, Cryptography, and Wireless Sensor Networks. 
	\end{IEEEbiography}

\end{document}